\documentclass[journal=jacsat,manuscript=article]{achemso}
\usepackage[version=3]{mhchem}
\usepackage{graphicx}
\usepackage{amsmath}
\usepackage{newtxmath}
\usepackage{textcomp}
\usepackage{xcolor}
\usepackage{soul}
\usepackage{array}
\usepackage[utf8]{inputenc}
\DeclareUnicodeCharacter{0308}{\"o}
\usepackage{fix-cm}
\usepackage{colortbl}
\usepackage{booktabs} 
\usepackage{multirow} 
\usepackage{units}
\usepackage{caption}
\usepackage{float}
\usepackage[normalem]{ulem}
\usepackage{hyperref}
\hypersetup{hidelinks,
    colorlinks=true,
    allcolors=black,
    pdfstartview=Fit,
    breaklinks=true}
\usepackage{verbatim}
\usepackage{lineno}
\usepackage{comment}
\usepackage{bm}
\makeatletter
\let\saved@includegraphics\includegraphics
\AtBeginDocument{\let\includegraphics\saved@includegraphics}
\renewenvironment{figure}
  {\@float{figure}}
  {\end@float}
\makeatother
\newcommand{\reftextit}[1]{}

\DeclareCaptionLabelFormat{adja-page}{\hrulefill\\#1 #2 \emph{(previous page)}}



\title[An \textsf{achemso} demo]
{SLICES-PLUS: A Crystal Representation Leveraging Spatial Symmetry}
\let\oldalign\align
\let\oldendalign\endalign
\renewenvironment{align}
{\linenomathNonumbers\oldalign}
{\oldendalign\endlinenomath}
\let\oldequation\equation
\let\oldendequation\endequation
\renewenvironment{equation}
{\linenomathNonumbers\oldequation}
{\oldendequation\endlinenomath}
\clearpage
\author{\normalsize Baoning Wang}
\affiliation[Tongji University]
{%
\footnotesize School of Physics Science and Engineering, Tongji University, Shanghai 200092, China}

\author{Zhiyuan Xu}
\affiliation[Tongji University]
{%
\footnotesize School of Physics Science and Engineering, Tongji University, Shanghai 200092, China}

\author{Zhiyu Han}
\affiliation[Tongji University]
{%
\footnotesize School of Physics Science and Engineering, Tongji University, Shanghai 200092, China}

\author{Qiwen Nie}
\affiliation[Tongji University]
{%
\footnotesize School of Physics Science and Engineering, Tongji University, Shanghai 200092, China}

\author{Hang Xiao}
\affiliation[Lingnan University]
{%
\footnotesize School of Interdisciplinary Studies, Lingnan University, Tuen Mun, Hong Kong SAR, China}
\email{hangxiao@ln.edu.hk}

\author{Gang Yan}
\affiliation[Tongji University]
{%
\footnotesize School of Physics Science and Engineering, Tongji University, Shanghai 200092, China}
\email{gyan@tongji.edu.cn}

\begin{document}
\maketitle
{\renewcommand{\baselinestretch}{1.7}
\clearpage 
\begin{abstract} 
In recent years, the realm of crystalline materials has witnessed a surge in the development of generative models, predominantly aimed at the inverse design of crystals with tailored physical properties. However, spatial symmetry, which serves as a significant inductive bias, is often not optimally harnessed in the design process. This oversight tends to result in crystals with lower symmetry, potentially limiting the practical applications of certain functional materials. To bridge this gap, we introduce SLICES-PLUS, an enhanced variant of SLICES that emphasizes spatial symmetry. Our experiments in classification and generation have shown that SLICES-PLUS exhibits greater sensitivity and robustness in learning crystal symmetries compared to the original SLICES. Furthermore, by integrating SLICES-PLUS with a customized MatterGPT model, we have demonstrated its exceptional capability to target specific physical properties and crystal systems with precision. Finally, we explore autoregressive generation towards multiple elastic properties in few-shot learning. Our research represents a significant step forward in the realm of computational materials discovery.

\end{abstract}
}
\maketitle
\renewcommand{\baselinestretch}{2}
\setlength{\parskip}{7pt}

\newpage
\section{\textbf{Introduction}}

\hspace{2ex} The application of artificial intelligence (AI) in materials science can be broadly categorized into two main paradigms: forward design and inverse design. Forward design refers to the prediction or characterization of material structures. Significant advances have been made in the domain of predictive modeling, with notable progress following the introduction of the CGCNN model\cite{xie2018crystal,yan2022periodic,chen2019graph,kaba2023prediction}.
Inverse design means fabricating or discovering novel materials with specific requirements. Historically, the discovery of new materials has largely depended on time-consuming trial-and-error approaches or serendipitous findings. Traditional approaches are resource-intensive and slow, which has limited the pace of material innovation. However, there has been a significant change in recent years. Advanced computational techniques in materials science have made it feasible to use deep learning for inverse design, a method that is becoming more widespread\cite{choudhary2022recent,vasudevan2019materials,mannan2024navigating,liu2023generative,liu2017materials}.

In contrast to traditional predictive models that begin with established material structures, the development of deep-generative models capable of inversely designing desired crystal structures presents a significant and intricate challenge\cite{noh2020machine}. The current landscape of mainstream inverse design methodologies encompass the following architectures: Variational Autoencoders (VAEs)\cite{kingma2013auto,ren_invertible_2022,court20203,korolev2020machine} , Generative Adversarial Networks (GANs)\cite{goodfellow2014generative,sanchez2018inverse,dan2020generative,long2021constrained} , diffusion models\cite{ho2020denoising}, and Crystal Diffusion Variational Autoencoders (CDVAEs)\cite{ye2024concdvaemethodconditionalgeneration,xie2022crystaldiffusionvariationalautoencoder} , among others.
Furthermore, the remarkable success of large language models (LLMs) in the fields of protein and drug design\cite{wang2024multi,ferruz2022protgpt2,lv2024prollama,wang2024token} has provided novel inspirations for inverse design of crystalline materials. This has led to the development of a series of large-language models dedicated to crystallography, including AtomGPT, CrystalFormer, MatterGPT, etc.\cite{choudharyexploring,cao2024space,chen2024mattergptgenerativetransformermultiproperty,wei2024crystal}. 
 
When examining materials generated through various methods, the focus is often on key attributes such as composition, properties, and structure configuration. Among these, spatial symmetry\cite{hiller1986crystallography} emerges as a crucial characteristic\cite{chen2022topological,tang2019comprehensive}. The presence of various symmetry types exerts significant constraints on the crystalline structures, profoundly influencing their practical application potential. For example, materials with high symmetry are particularly valued in the fields of solar energy batteries and mechanical engineering\cite{cao2017cubic,luan2018mechanical,mouhat2014necessary}. Their structural stability and uniformity endow them with exceptional performance characteristics. Materials with low symmetry, on the other hand, may have advantages in the fabrication of optoelectronic devices and the investigation of surface plasmon phenomena. Their anisotropic physical properties enable special responses to external stimuli, making them indispensable in these applications\cite{hong2021n,zheng20242d,tan2024deformable,meng2016integration}.

However, existing works tend to overlook the significant constraints of space group symmetry\cite{chenebuah2024deep,jiao2024space}, leading to the generation of materials with low symmetry, such as triclinic or monoclinic crystal structures. Aware of this, some researchers have succeeded in promoting the effectiveness and reliability of crystal formation by considering the effects of symmetry\cite{chenebuah2024deep,gruver2024fine,zhao2023physics,zhu2023wycryst}, yet these approaches essentially involve random generation and require extensive screening to identify crystals that are consistent with specific criteria. Furthermore, there are efforts to directly design crystals with target crystal systems or space groups\cite{zhao2021high,cao2024space,jiao2024space}. But these methods are restricted by either limited diversity or a lack of targeted directionality, and fail to modulate physical properties of materials simultaneously.

Hence, there is a pressing need to develop an efficient method for the target generation of materials with specific symmetric structures and desired physical properties. Drawing inspiration from the Simplified Molecular Input Line-Entry System (SMILES) for molecular representation\cite{weininger1988smiles}, Xiao et al. have introduced an innovative and invertible crystal representation method known as simplified line-input crystal-encoding system (SLICES)\cite{xiao2023invertible}. It stands out as a leading approach in crystal structure representation, boasting exceptional interpretability and scalability. SLICES features a straightforward and concise syntax, with a capacity for easy expansion of its functionality. The string-based representation of crystal structures, in conjunction with the advanced MatterGPT model, facilitates efficient model training and enables precise generative analytics\cite{chen2024mattergptgenerativetransformermultiproperty}.

SLICES has shown promise in the targeted design of crystals, focusing on properties such as formation energy and band gap\cite{xiao2023invertible,chen2024mattergptgenerativetransformermultiproperty}. However, its application in symmetry-guided crystal generation has not been thoroughly investigated. To bridge this gap, we introduce SLICES-PLUS, an advanced variant of SLICES that maintains its core benefits and enhances the description of complex spatial symmetries in crystal structures. Through our comparative experiments, we have demonstrated that SLICES-PLUS outperforms the original SLICES in improving the structural validity of the generated crystals under identical conditions. Additionally, we have proven SLICES-PLUS's effectiveness in generating crystals with desired property and space group simultaneously.
We have also applied SLICES-PLUS to few-shot learning, illustrating its advances at addressing target multiple properties while retaining its focus on symmetry features. This research provides valuable insights into the utility of SLICES for material generation and offers constructive advancements for its application in various task scenarios.

\section{\textbf{Results and Discussions}}
\subsection{Design of SLICES-PLUS}
\hspace{2ex} Three-dimensional crystals are organized into 230 distinct space groups and fall into 7 different crystal systems. This categorization is fundamentally based on unique symmetry characteristics of each space group. The symmetry of each space group encompasses two primary elements: spatial rotational symmetry and translational symmetry.
Rotational symmetry within a space group can be mathematically depicted by 3-dimensional rotation matrices, which describe the angles and axes of rotation. Meanwhile, translational symmetry is conveyed through translation vectors expressed in fractional coordinates, indicating how the crystal pattern repeats periodically in space. For instance
\begin{equation}
    \begin{aligned}
        \mathbf{r}' &= \{\mathbf{R}|\tau\} \mathbf{r} = \mathbf{R}\mathbf{r} + \tau \\
        &= \begin{pmatrix}
            1 & 0 & 0 \\
            0 & -1 & 0 \\
            0 & 0 & 1
        \end{pmatrix} \begin{pmatrix}
            x \\
            y \\
            z
        \end{pmatrix} + \begin{pmatrix}
            1/2 \\
            1/2 \\
            0
        \end{pmatrix} = \begin{pmatrix}
            x + 1/2 \\
            -y + 1/2 \\
            z
        \end{pmatrix}
    \end{aligned}
\end{equation}
If multiple atoms within a unit cell share identical spatial position information, they are considered to have the same symmetry operation. It implies that they can be transformed into each other through this symmetry operation. Consequently, a greater number of shared symmetry operations corresponds to a higher degree of symmetry in crystal structures.

Wyckoff Positions are a system used to describe the symmetry of valence atoms in unit cells comprehensively. Each Wyckoff position corresponds to a set of points that share the same symmetry operations. The depictive notation progresses from special positions at the bottom to general positions at the top with multiplicity increasing. In formula (1), the general wyckoff position could be expressed as (x+1/2, -y+1/2, z). These systematic Wyckoff positions enable us to precisely identify and differentiate the symmetries among 230 space groups. To simplify our representations, we focus on the general Wyckoff positions in this study, which are all available from the symmetrized CIF files.
\begin{figure}[htbp]
    \centering
    \includegraphics[width=16cm]{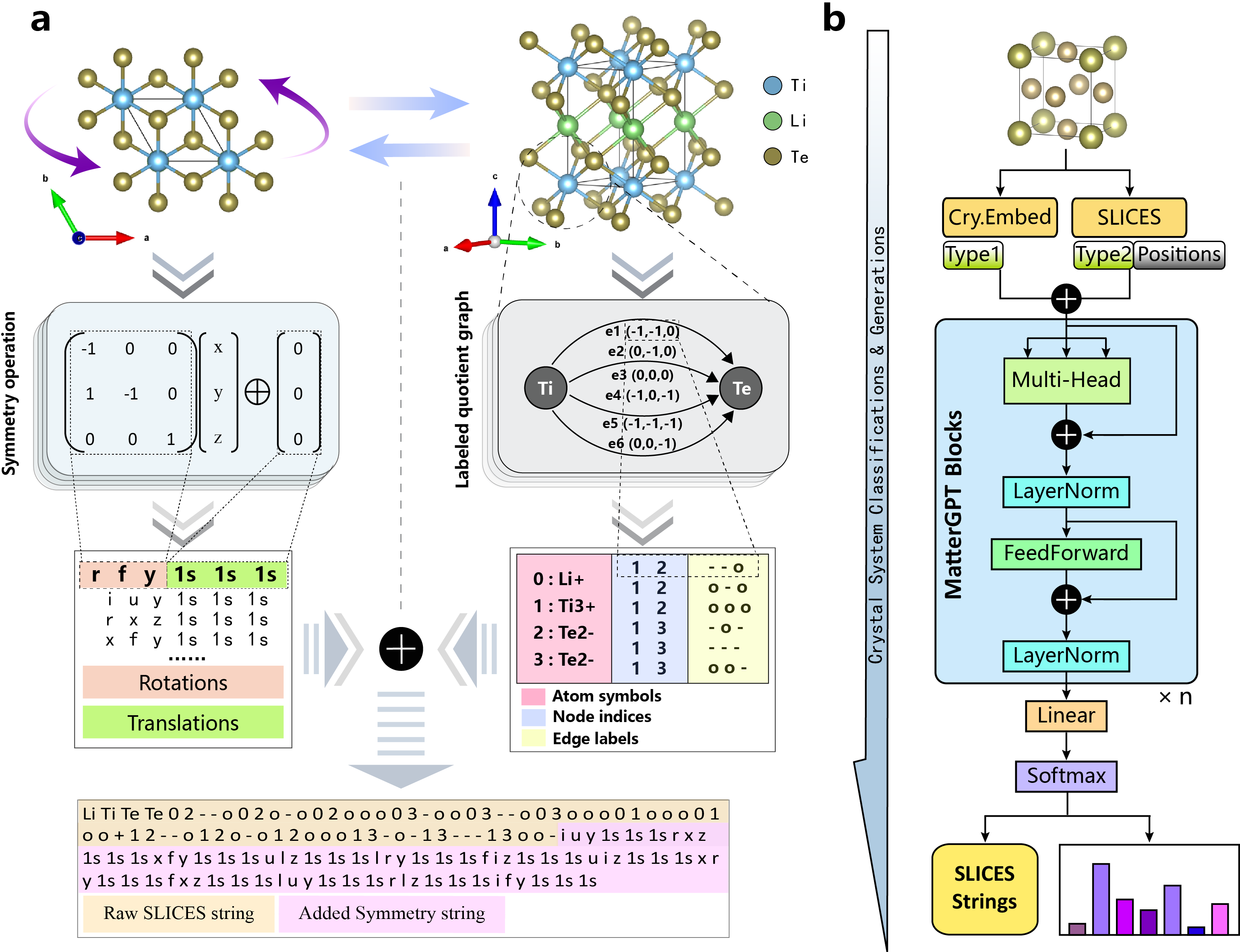}
    \caption{\textbf{Composition of SLICES-PLUS. (a) }Just as the label quotient graph can serve as an intermediary to translate between the crystal structure and the SLICES strings, by encoding symmetry operations (translation, rotation, inversion) into strings, we get a symmetry sequence. These two parts add up to form SLICES-PLUS. \textbf{(b) }Pipline for classification and generation tasks using the conditional MatterGPT model.
    }
    \label{fig:fig1}
\end{figure}

Figure 1a outlines the design concept of SLICES-PLUS. Building upon the foundation of SLICES, we have integrated the spatial symmetry depicted of the Wyckoff notation. In detail, we separate the symmetry operations in crystal structures into two components: rotation matrices and translation vectors. Rotation matrices exhibit distinctive unitary attributes, with each element being restricted to the values of -1, 0, or 1. Which implies that the total number of potential configurations for every row or column is 3 $\times$ 3 = 27, yet rows or columns that are entirely zero are not allowed, thus narrowing the possibilities to 26. By association with the alphabet, we can use the 26 letters to represent each row of the rotation matrix in a straightforward manner. The translation vector, meanwhile, is expressed as three-dimensional fractional coordinates relative to the unit cell dimensions. The physical constraints of solid crystals allow only for 1-, 2-, 3-, 4-, and 6-fold rotation axes or rotoinversion axes. These constraints are mirrored in the translation vector, which only incorporates the fractional coordinates of 0, 1, $\frac{1}{2}$, $\frac{1}{3}$, $\frac{2}{3}$, $\frac{1}{4}$, $\frac{3}{4}$, $\frac{1}{6}$ and $\frac{5}{6}$. So we can utilize a straightforward combination of Arabic numerals and letter`s' to denote the fractional coordinates of the translation vector, such as `1s', `2s', `3s' et al. 

By merging the rotation matrix and translation vector information, any symmetry operation can be encoded in a concise 6-character string, akin to `r,f,y,1s,1s,1s'. This concatenation of codes unpacks a wealth of three-dimensional structural information within the crystals, which is then appended to the original SLICES, thereby augmenting the depictive strength of SLICES-PLUS. Note that the original SLICES remains inherently self-consistent with the newly integrated symmetry sequences, ensuring that the inclusion of symmetry information does not interfere with the decoding process of the existing SLICES. Thus highlights the extensibility of SLICES for crystal encoding.

\subsection{Comparison between SLICES-PLUS and SLICES}
\hspace{2ex} To showcase the benefits of SLICES enhanced with symmetry strings, we train a MatterGPT model using SLICES-PLUS for crystal system-based classification and generation tasks, and compare with the original SLICES. In all of the following experiments, we assign the seven crystal systems to numerals ranging from 0 to 6. Specifically, the numbers 0 through 6 denote cubic, hexagonal, trigonal, tetragonal, orthorhombic, monoclinic, and triclinic systems, respectively.

The dataset used in classification experiment is 2000 samples per crystal system, and the space groups in each crystal system are randomly distributed. To assess the impact of symmetry encoding on the precision of crystal system classification, we performed a five- and a three-class classification task. The results were described through confusion matrices, as illustrated in Figure 2. Panels 2(a) and 2(b) display the prediction results using SLICES, while panels 2(c) and 2(d) exhibit the results with SLICES-PLUS. It is evident that the incorporation of three-dimensional symmetry significantly enhances the predictive accuracy for all crystal systems. Except for cubic system, the prediction accuracies for the other systems using SLICES are approximately from 80$\%$ to 90$\%$, whereas with SLICES-PLUS, they all exceed 90$\%$ statistically. It indicates that SLICES-PLUS indeed encompasses more spatial information than its predecessor.

\begin{figure}[htbp]
    \centering
    \includegraphics[width=16cm]{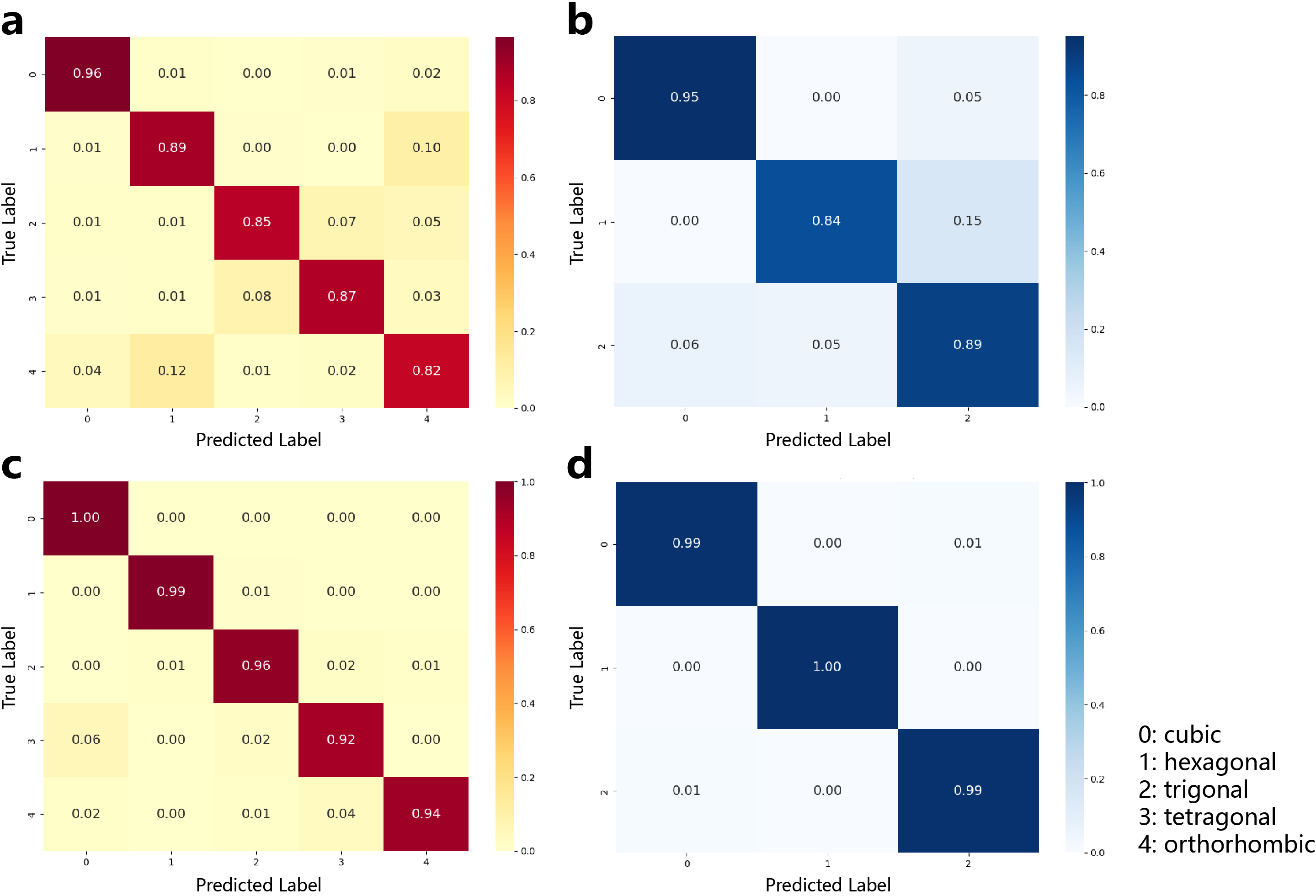}
    \caption{\textbf{Crystal system based classification tasks using SLICES and SLICES-PLUS representation (cubic, hexagonal, trigonal, tetragonal, orthorhombic). (a,c) }Five categories and \textbf{(b,d) }Three categories.
    }
    \label{fig:fig2}
\end{figure}
Next, we conducted a generation task, the model was guided purely by the crystal system types, and 500 sequences are sampled for each system after training. The strings sampled were then reconstructed into crystal structures, with only the valid ones being filtered and displayed in the bar chart in Figure 3. For this generation process, we focused on four target crystal systems: cubic, hexagonal, trigonal, and tetragonal. The blue bars in the chart represent the statistics of valid crystals generated using SLICES-PLUS, while the brown bars indicate those generated with the original SLICES. It is our initial attempt to apply MatterGPT to categorical generation tasks. Obviously, the symmetry properties of different space groups are different even in the same crystal system, which would cause disturbance to each other. In order to reduce the distraction, for each type of crystal system in the training set and test set, we focus on single typical space group to represent the geometric features (Supplementary Table S2). 
\begin{figure}[htbp]
    \centering
    \includegraphics[width=16cm]{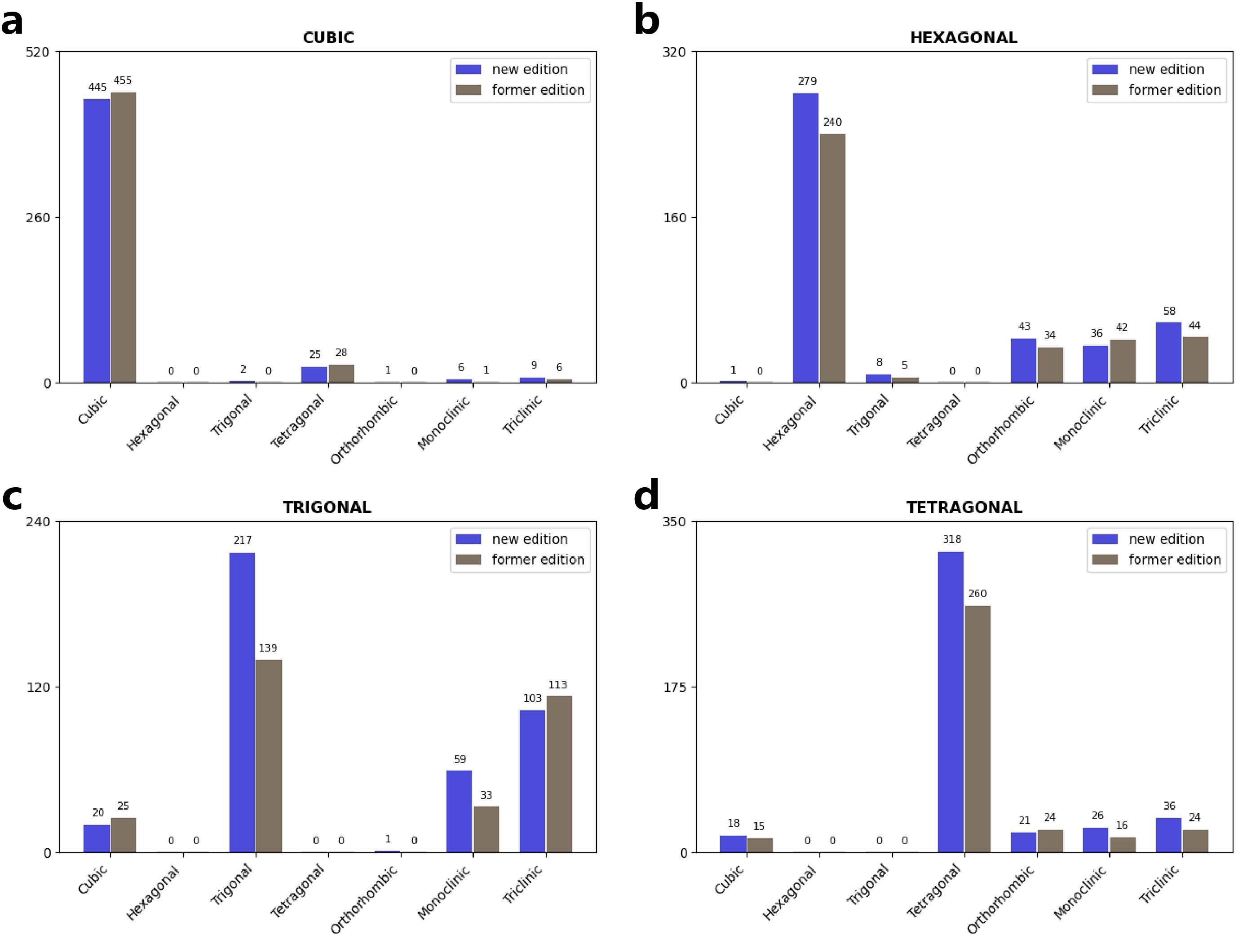}
    \caption{\textbf{Crystal system based generation tasks using SLICES and SLICES-PLUS representation. }The distributions of crystals sampled with target system of \textbf{(a) }cubic \textbf{(b) }hexagonal \textbf{(c) }tetragonal \textbf{(d) }trigonal.}
    \label{fig:fig3}
    \vspace{1em} 
\end{figure}
\begin{table}[htbp]
    \centering
    \captionsetup{labelfont=bf}
    \caption{Generative Performance for Several Target Systems}
    \setlength{\tabcolsep}{4pt} 
    \resizebox{\textwidth}{!}{ 
    \fontsize{13}{15}\selectfont 
    \begin{tabular}{p{30mm}>{\centering\arraybackslash}p{35mm}>{\centering\arraybackslash}p{40mm}>{\centering\arraybackslash}p{40mm}>{\centering\arraybackslash}p{40mm}}
        \toprule[0.8pt]
        \toprule[0.8pt]     
        Representation & Target & Validity ($\%$) & Uniqueness ($\%$) & Novelty ($\%$) \\
        \cmidrule[0.6pt]{1-5}
        \multirow{4}{*}{SLICES} & Cubic & 98.2 & 92.0 & 45.4 \\
        & Hexagonal & 73.8 & 97.2 & 58.5 \\
        & Trigonal & 65.6 & 96.6 & 70.8 \\
        & Tetragonal & 67.8 & 95.2 & 57.2 \\
        \cmidrule[0.6pt]{1-5}
        \multirow{4}{*}{SLICES-PLUS} & Cubic & 97.6 & 94.0 & 48.4 \\
        & Hexagonal & 85.6 & 96.8 & 56.4 \\
        & Trigonal & 80.8 & 96.2 & 66.5 \\
        & Tetragonal & 84.0 & 92.6 & 57.1 \\
        \bottomrule[0.8pt]
        \bottomrule[0.8pt]     
    \end{tabular}}
    \vspace{-2mm}
    \label{tab:merged_rows}
\end{table}
Comparing the two types of data distributions, it is clear that except for cubic crystal samples(both own validity over 95$\%$), the validity of other crystal samples expressed by SLICES-PLUS are more than 10 percentage points higher than those expressed by SLICES stably. In addition, the proportion of target crystal systems expressed in SLICES-PLUS is also higher than that of the former, suggesting that the model captures the decisive characteristics of each system more effectively. Other indicators, such as novelty and uniqueness, are not much different.

\begin{figure}[htbp]
    \centering
    \includegraphics[width=16cm]{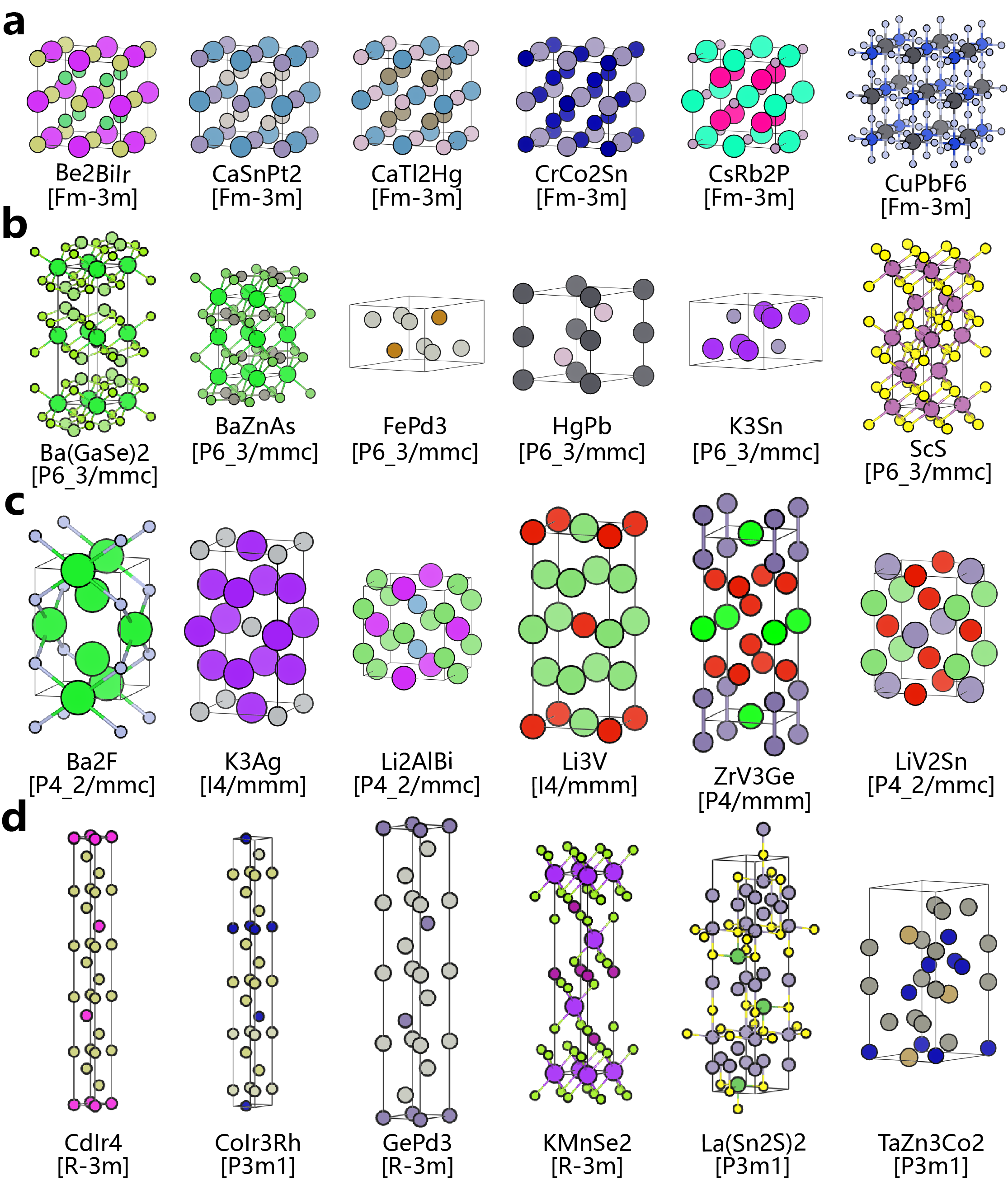}
    \caption{\textbf{Some of the novel crystals discovered by our model} with crystal system of \textbf{(a)} Cubic, \textbf{(b)} Hexagonal, \textbf{(c)} Tetragonal, \textbf{(d)} Trigonal.}
    \label{fig:fig4}
\end{figure}
\subsection{Conditional Generation based on Crystal Symmetry and Physical Property}
\hspace{2ex} Generating crystals with target spatial symmetry alone may not be sufficient. The pivotal point lies in our ability to maintain the generation efficiency of desired physical properties while simultaneously controlling the spatial features. For example, sometimes in semiconductor or battery design, there is a preference for materials that not only have appropriate electric property but also exhibit high symmetry\cite{wei2023triggering,he2024leveraging}. Therefore, we introduce a dual attention mechanism that leverages both spatial symmetry and numerical properties for the conditional generation of crystal structures. This architecture employs two distinct workflows to facilitate a generation process that is guided by crystal systems and additional properties, which is formation energy (can also be predicted by M3GNet\cite{chen2022universal} to speed up) in our demo. 

As depicted in Figure 4(a), when generation is conditioned on a cubic crystal system and a formation energy of -0.5eV, the resulting crystal structures tend to cluster around these specific parameters. Figure 4(b) similarly demonstrates that when the conditions are hexagonal crystal system and formation energy of -1.5eV, the generated structures align closely with these criteria. The rationale behind utilizing two workflows is to emphasize the regression of physical properties while also incorporating the classification of the crystal systems, ensuring that these processes do not interfere with each other. Our dual workflow model, in conjunction with SLICES-PLUS, achieves a high level of generation performance. The results indicate that our crystal generation approach not only efficiently steers the creation of specific symmetric structures but also adeptly manages the physical properties of the resulting crystals.

\begin{figure}[htbp]
    \centering
    \includegraphics[width=16cm]{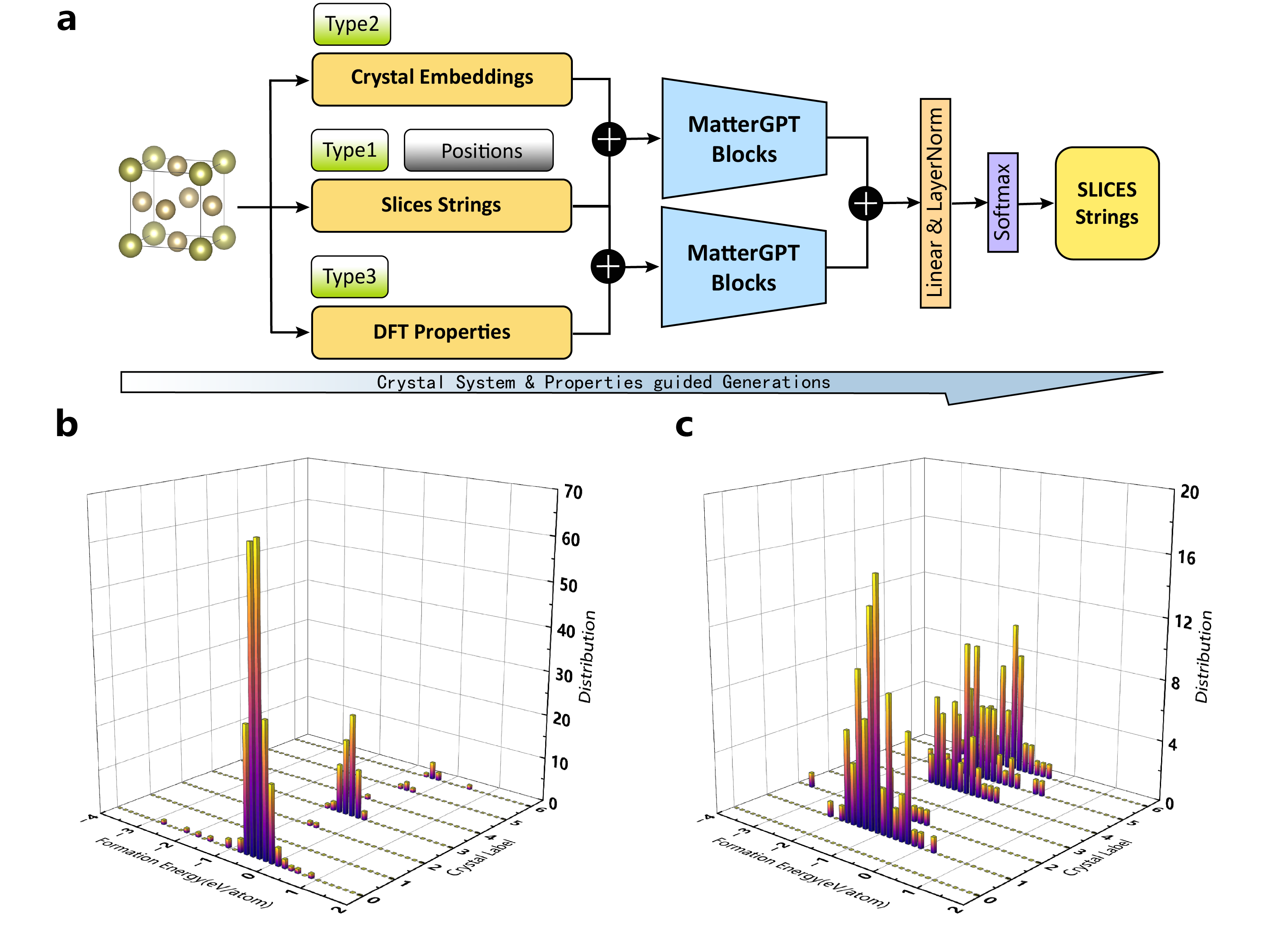}
    \caption{\textbf{De novo crystal generation targeting specific formation energy and crystal system. (a) }Customized dual MatterGPT workflows for space-property generation. \textbf{(b,c) }distribution of sampled crystals targeting $\{$$E_f$: -0.5 eV/atom, crystal label: 0(cubic) $\}$ and $\{$$E_f$: -1.5 eV/atom, crystal label: 1(hexagonal)$\}$}.
    \label{fig:fig5}
\end{figure}

\subsection{Conditional Generation for Cubic crystals based on multiple elastic conditions}
\hspace{2ex} The experiments conducted have highlighted the superiority of SLICES-PLUS in the generation of high-symmetry crystal structures. This advantage is attributed to the enhanced spatial features that these crystals exhibit during the symmetry encoding process, which facilitate their identification and replication by the model. In our subsequent explorations, we opt to forgo comparisons across various crystal systems and instead concentrate our efforts on the meticulous construction of cubic crystals, thereby maximizing the potential of SLICES-PLUS in this specialized domain.

In this phase of our research, we delve into the assessment of the elastic properties of crystals: bulk modulus (K) and shear modulus (G). Given the scarcity of existing elastic data, we meticulously filtered and screened to compile a sparse dataset comprising 4071 cubic crystals. Leveraging MatterGPT's strength in generating new crystals tailored to multiple property combinations, we concurrently learn the bulk modulus and shear modulus of these materials. To test the learning effect, we sample crystals for two different pairs of objective values for (K, G): (200 GPa, 100 GPa) and (50 GPa, 30 GPa).

The results reveal that for two target pairs, we successfully transform 295 and 294 samples into valid crystal structures, respectively. Note that despite the absence of explicit embedding layers for the features of crystal system in this model, as was the case in the previous three experiments, a significant proportion of the samples are correctly identified as cubic crystals. Specifically, there are 255 and 243 cubic crystals in the two sets, occupying 86.4$\%$ and 82.7$\%$ of the valid samples, respectively.

In addition, the density distribution maps presented in Figures 5(a) and 5(b) clearly illustrate that the property values of the generated samples predominantly cluster around our predefined target values. Additionally, we have depicted the three-dimensional distribution of the three key variables within the elastic tensor of cubic crystals. These results substantiate the proficiency of SLICES-PLUS in achieving target generation of high-symmetry crystals with multiple desired properties, even when operating on a sparse dataset.
\begin{figure}[H]
    \centering
    \includegraphics[width=16cm]{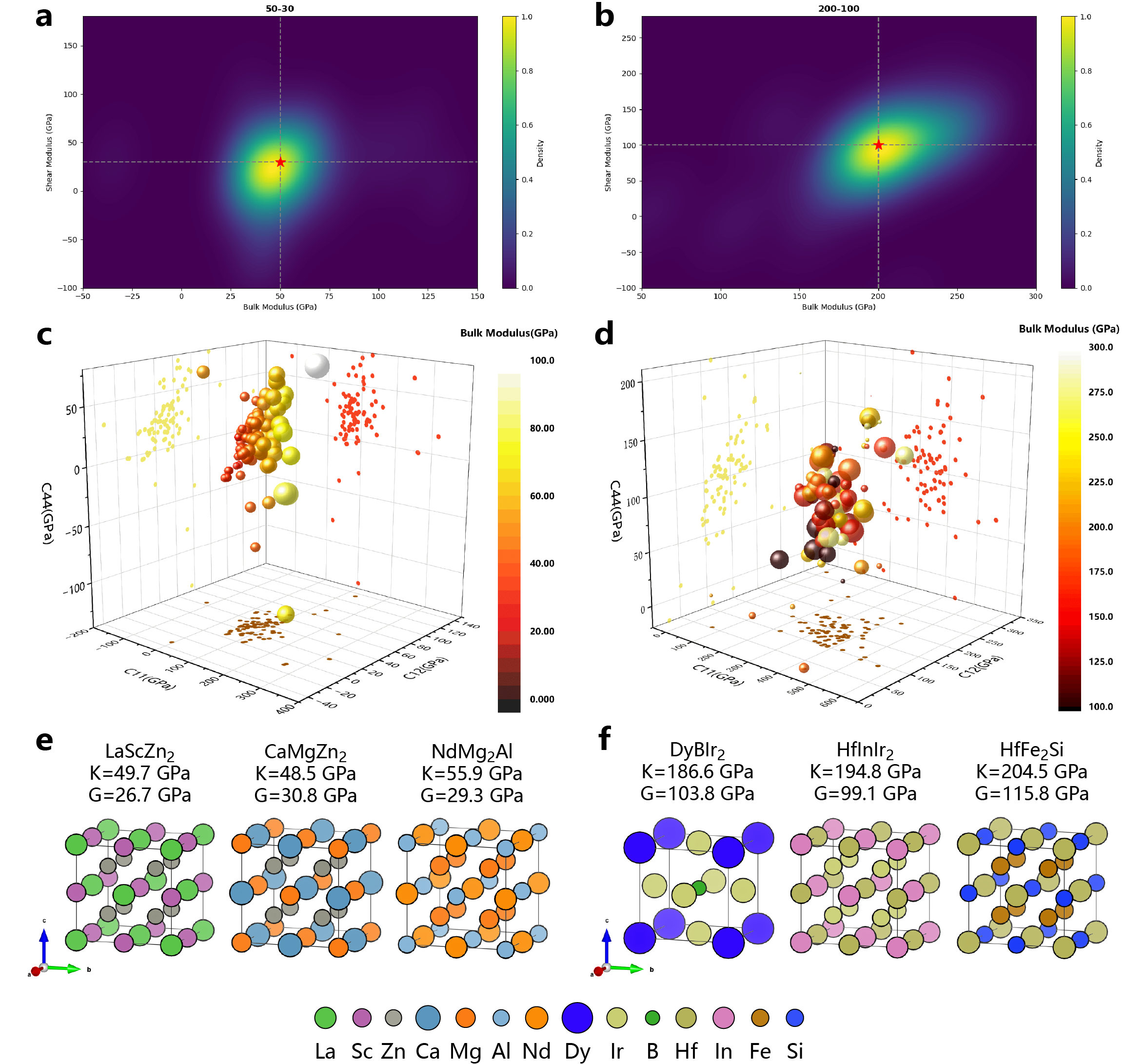}
    \caption{\textbf{De novo crystal generation targeting desired (K, G) pairs based on few-shot learning. (a,b) }distribution of sampled crystals aiming $\{$$K$: 50GPa, $G$: 30GPa $\}$ and $\{$$K$: 200GPa, $G$: 100GPa$\}$. \textbf{(c,d) }3D distribution of sampled crystals with independent elastic tensor variables as coordinates. \textbf{(e) }standard orientation shapes of 3 novel de novo generated crystals with desired ($K$,$G$).
    }
    \label{fig:fig6}
\end{figure}

\section*{\textbf{Conclusions and Outlooks}}
\hspace{2ex} In summary, we have developed a variant of SLICES called SLICES-PLUS and demonstrated its advantages for spatial symmetry-based generation tasks over the original edition. Our research underscores the scalability and adaptability of SLICES. Furthermore, in combination with a customized dual pipeline MatterGPT, we have used SLICES-PLUS to generate crystal materials that target specific spatial symmetry and physical properties simultaneously. Finally, we apply SLICES-PLUS to few-shot learning and indicate its strength to generate cubic crystals targeting multiple properties. 

However, as a dilemma faced by most data-driven deep learning approaches, the scarcity of high-quality computational material samples somewhat constrains the potential of SLICES. Although SLICES-PLUS has shown superior convergence and robustness over SLICES in symmetry-based target generation tasks (Supplementary Figure S1), making it a more attractive option when dealing with limited and diverse data samples, there is still room for improvement. Given these insights, future improvements could potentially be realized by further refining the MatterGPT architecture and the method of embedding information, thereby enhancing the efficiency and accuracy of crystal generation. We can also introduce more physical prior knowledge to develop a half-data, half-principle-driven intelligent model.

\section*{\textbf{Methods and Materials}}
\subsection{SLICES-PLUS representation}
\hspace{2ex} SLICES represents a pioneering approach in crystallography, being the first reversible and invariant representation of crystals. It is made up of three fundamental elements: atomic symbols, node indices, and edge labels. Building upon this foundation, we have introduced novel fundamental units in SLICES-PLUS, which include row symbols of the rotation matrix and coordinate symbols of the translation vector. This enhancement enriches the encoding scheme, providing a more comprehensive framework for crystal structure representation. As highlighted previously, the finite and distinct nature of these basic unit elements facilitates straightforward labeling and dictionary mapping during the training process. By amalgamating these fundamental units, we can derive a comprehensive SLICES-PLUS representation sequence.

The symmetry sequence encoding method is described as follows. We utilize Structure from Pymatgen to parse CIF files and extract structural data. Subsequently, we employ the SpacegroupAnalyzer class to conduct space group analysis on the obtained structure. Finally, we extract the symmetry operations, encapsulated by rotation matrices and translation vectors as indicated by Wyckoff positions, to formulate symmetry strings. This approach ensures robust and precise encoding of the crystal symmetry within SLICES-PLUS.

The SLICES-PLUS decoding process adheres to the same principles as SLICES, employing the SLI2Cry algorithm for this purpose. Given that the original SLICES fundamentally represents the crystal graph's adjacency matrix, and the symmetry sequence essentially unveils the latent spatial attributes of the crystal, these two pieces of information are inherently self-consistent, allowing the existing decoding method to be seamlessly extended.

SLICES-PLUS not only maintains the high reversibility and invariance of the original one but also serves to direct the model's attention towards the three-dimensional spatial characteristics of the crystal. This improvement bolsters the model's capacity to discern and generate target crystal structures accurately.
\subsection{Dataset construction}
\hspace{2ex} The entire dataset for training and testing in this work are sourced from the Materials Project\cite{blochl1994projector}. In order to ensure the effectiveness of the training process, it is crucial to filtering datas based on the crystal system characteristics during the construction. The refined methodology is described below.

1. Utilize MPRester and the associated API to find the structure of the target crystal;\\
\indent 2. Use CifWriter to generate CIF files, setting the symprec parameter to 0.1 to accommodate the positional deviations permitted for symmetry identification;\\
\indent 3. Execute the SLICES-PLUS encoding and decoding process to ensure the integrity of the representation;\\
\indent 4. Analyze the crystal system of the decoded crystal to verify its accuracy.

Any instances where the decoded crystal is found to belong to a crystal system different from the original one are flagged as problematic and are subsequently excluded from the dataset. This rigorous approach ensures the reliability of the data set and the model's ability to accurately interpret and generate crystal structures.

Further refining our dataset, we have applied additional steps: focusing on materials with fewer than 35 atoms per unit cell, excluding atoms beyond atomic number 86, and disregarding low-dimensional structures like molecular crystals. This process has yielded a high-quality dataset that includes a variety of crystal structures, their numerical properties, and crystal system information, ensuring the validity of the dataset for model training and testing. Distributive details of the dataset for Figure 3 and Figure 4 are available in Supplementary Information.

Given the scarcity of elastic property data, the dataset for the bulk-shear experiment shown in Figure 5 only comprise 4071 cubic crystals and excluding any data exhibiting markedly abnormal values. The selected data exhibit bulk modulus values that range predominantly from 0 GPa to 300 GPa, and the shear modulus values are largely concentrated between -200 GPa and 200 GPa. Further details on the distribution of datasets are available in Supplementary Figure S3.

\subsection{Architecture of model}
\hspace{2ex} Our model is a custom-trained version of MatterGPT, designed for symmetry-based crystal generation tasks. As a transformer decoder, MatterGPT is designed to generate solid materials from scratch. When integrated with SLICES-PLUS, it effectively captures the syntactic patterns and structural nuances of the crystals.

We add a CrystalEmbedding layer, which supplements the standard tokenizer and positional encoding. This layer translates the labels of each crystal system into a consistent vector format. Notably, the inclusion of this layer has been shown to significantly boost the model's performance, and it also serves as a key conditional element for autoregressive generation.

The primary structural distinction between Experiment 1 and Experiment 2 lies in the configuration of the output layer dimensions: Experiment 1 uses a dimension of $num\_classes$, which is geared toward classifying crystal systems, while Experiment 2 employs a dimension of $voc\_size$, intended for the generation of crystal systems.

For Experiment 3, the model receives inputs for SLICES-PLUS(crystal graph), property sequence, and crystal label sequence as detailed below:
\begin{equation}
R_{slices} = Drop[X_{slices};X_{type}^{1};X_{position};X_{crystal}]
\end{equation}
\begin{equation}
R_{property} = [X_{property};X_{type}^{2}]
\end{equation}
\begin{equation}
R_{crystal} = [X_{crystal};X_{type}^{3}]
\end{equation}
Where $X_{slices}$ is the SLICES-PLUS sequence, $X_{position}$, $X_{type}^{1,2,3}$, $X_{crystal}$ and $X_{property}$ refer to the positional encoding, sequence type, crystal label, and the numerical property mapping, respectively. $Drop$ is a dropout layer used to prevent overfitting. $[*;\ldots;*]$ denotes the concatenation operation of several tensors.

In Experiment 3, we adopted a dual workflow to accommodate both the regression generation of continuously distributed DFT properties and the classification generation of discretely distributed crystal systems. This means that each is independently trained through separate MatterGPT-Blocks, effectively applying self-attention mechanisms to spatial symmetry structures and numerical properties, respectively.
\begin{equation}
R_{crystal}^{'} = MatterBlocks_1(R_{slices} \oplus R_{crystal})
\end{equation}
\begin{equation}
R_{property}^{'} = MatterBlocks_2(R_{slices} \oplus R_{property})
\end{equation}
Where $\oplus$ denotes the concatenation operation of two inputs. $MatterBlocks_{1,2}$ contain multi-head attention layers, feed-forward layers, and residual layers. The structure of a single Block is as follows:
\begin{equation}
    MatterBlock(x) = Linear[\Phi(q)] + q
    \end{equation}
\begin{equation}
q = MultiAttention[\Phi(x)] + x
\end{equation}
Where $\Phi$ denotes the layer normalization operation. Then, we concatenate the outputs of the two workflows, integrating the features captured by both paths. It is followed by a linear layer that maps the tensor back to the hidden layer size.
\begin{equation}
R_f = \Phi[Linear(R_{crystal}^{'} \oplus R_{property}^{'})]
\end{equation}
Finally, cross-entropy is used to obtain the desired output results. The use of dual workflows largely avoids information interference in multitask learning, improving the generation ability. Combined with the Space-Properties dual attention mechanism, the model performs well in generating tasks that consider both spatial symmetry and numerical properties. The model parameter settings used in this work are detailed in Supplementary Table S1.
\subsection{DFT calculation and Evalution method}
\hspace{2ex} We use the Vienna Ab initio Simulation Package (VASP)\cite{kresse1996efficiency} to perform DFT calculations on the generated crystal samples, employing the PAW method\cite{blochl1994projector} and the Perdew-Burke-Ernzerhof(PBE) functional\cite{perdew1996generalized}. The energy convergence and force convergence criteria are set to be compatible with the MP database.

Before formal calculations, we uniformly use MPRelaxSet\cite{ong2013python} for structural relaxation to ensure the precision of the calculation and speed up the calculation. For calculations of the formation energy, we first obtain the total energy($E_{total}$) of materials, and then calculate the average energy of each atom($i$) using the following equation. 
\begin{equation}
    E_f = \frac{E_{total} - \sum{n_{i}E_{i}}}{\sum{n_i}}
\end{equation}
For bulk modulus and shear modulus calculations, we obtain the elastic tensor data after relaxation by setting special INCAR parameters. Elastic tensor is a 6-dimensional diagonal matrix, which generally has 21 independent variables (triclinic system). However, for the cubic system in which we are interested, there are only 3 independent matrix elements ($C_{11}$, $C_{12}$, $C_{44}$).
\begin{equation}
    \begin{pmatrix}
        C_{11} & C_{12} & C_{12} & 0 & 0 & 0 \\
        C_{12} & C_{11} & C_{12} & 0 & 0 & 0 \\
        C_{12} & C_{12} & C_{11} & 0 & 0 & 0 \\
        0 & 0 & 0 & C_{44} & 0 & 0 \\
        0 & 0 & 0 & 0 & C_{44} & 0 \\
        0 & 0 & 0 & 0 & 0 & C_{44}
    \end{pmatrix}
\end{equation}
Based on the elastic constants obtained from the VASP calculations, we can derive the formulas for the average shear modulus ($G$) and the average bulk modulus ($K$) using the Voigt-Reuss-Hill\cite{voigt1928lehrbuch,reuss1929calculation,hill1952elastic} approximation model. Taking cubic crystals as an example:
\begin{equation}
G = \frac{1}{2}(G_V+G_R) = \frac{1}{10}[(C_{11}-C_{12})+3C_{44}] + \frac{5C_{44}(C_{11}-C_{12})}{6(C_{11}-C_{12})+8C_{44}}
\end{equation}
\begin{equation}
K = \frac{1}{2}(K_V+K_R) = \frac{1}{3}(C_{11}+2C_{12})
\end{equation}
Where $G_V$ and $K_V$ are the Voigt estimates, which assume that the components of the composite material are in a state of equal stress but different strain, usually representing the upper bound estimate of the mechanical modulus. $G_R$ and $K_R$ are the Reuss estimates, which assume that the components are in a state of equal strain but different stress, usually representing the lower bound estimate of the mechanical modulus.

All of the mechanical information can be directly extracted using the ElasticTensor utility in Pymatgen\cite{ong2013python}. Finally, we plotted the calculated DFT values as a two-dimensional density map to observe their distribution around the target points, verifying the learning effectiveness.

In addition, to evaluate the performance of the crystal generation model, we used validity, uniqueness, and novelty as criteria. Validity refers to the percentage of generated strings that conform to the SLICES-PLUS representation syntax and can be reconstructed into crystal structures. Uniqueness refers to the percentage of nonredundant samples among valid samples. Novelty refers to the percentage of valid crystals that do not exist in the training set, calculated using the StructureMatcher utility implemented in Pymetgen\cite{ong2013python}.

\section*{\textbf{Data availability}} 
The source data and code are available on Github (\href{https://github.com/wbnbb/SLICES-PLUS}{https://github.com/wbnbb/SLICES-PLUS}).

\section*{\textbf{Acknowledgements}} 
We thank X.Ru, X.Wu, Z.Wang and Y.Xu for helpful discussion. We thank S. Lu for his assistance in applying DFT calculations.

\section*{\textbf{Author Contributions}}
B.W. conceived the idea and designed the work. H.X. and B.W. implemented the models. Z.X. and Z.H. performed the analysis. B.W.,Z.X.,Z.H.and Q.N. wrote the manuscript. G.Y. and H.X. contributed to the discussion and revision.

\section*{\textbf{Notes}}
The authors declare no competing financial interests.

    \bibliography{bb}

\newpage
\section*{\textbf{Supplementary Information}}
\begin{table*}[htbp]  
    \caption*{\textbf{Table S1}: The hyperparameters of MatterGPT model}
    \centering
    \setlength{\tabcolsep}{4pt} 
    \large 
    \begin{tabular}{p{100mm}>{\centering\arraybackslash}p{40mm}}
        \toprule[0.8pt]
        \toprule[0.8pt]     
        Optimizer & Adam \\
        Learning rate & 0.0005 \\
        Batch size & 10 \\
        Embedding dropout & 0.1 \\
        Attention dropout & 0.1 \\
        Output projection dropout & 0.1 \\
        Embedding size & 512 \\
        Feedforward size & 2048 \\
        Attention heads & 8 \\
        Number of layers & 8 \\
        \bottomrule[0.8pt]
        \bottomrule[0.8pt]     
    \end{tabular}
    \vspace{-2mm}
    \label{tab:mattergpt_hyperparameters}
\end{table*}

\newpage
\begin{table*}[htbp]  
    \caption*{\textbf{Table S2}: Training samples count the number of samples in the training set(Figure3). Multiplicity represents the number of equivalent positions in the space group. Fractional coordinates list the coordinates of the equivalent positions in the space group.}
    \centering
    \setlength{\tabcolsep}{4pt} 
    \resizebox{\textwidth}{!}{ 
    \large 
    \begin{tabular}{p{20mm}>{\centering\arraybackslash}p{25mm}>{\centering\arraybackslash}p{20mm}>{\centering\arraybackslash}p{20mm}>{\centering\arraybackslash}p{90mm}}
        \toprule[0.8pt]
        \toprule[0.8pt]     
        Space group & Crystal system & Training samples & Multiplicity & General Wyckoff positions \\
        \cmidrule[0.6pt]{1-5}
        255 & Cubic & 1785 & 192 & (x, y, z), (-x, -y, z), (-x, y, -z), (x, -y, -z), (z, x, y), (z, -x, -y), (-z, -x, y), (-z, x, -y), (y, z, x), (-y, z, -x), (y, -z, -x), (-y, -z, x), …… (-x+1/2, -z+1/2, y), (x+1/2, -z+1/2, -y), (x+1/2, z+1/2, y), (-x+1/2, z+1/2, -y), (-z+1/2, -y+1/2, x), (-z+1/2, y+1/2, -x), (z+1/2, -y+1/2, -x), (z+1/2, y+1/2, x) \\
        
        \cmidrule[0.6pt]{1-5}
        194 & Hexagonal & 1971 & 24 & (x, y, z), (-y, x-y, z), (-x+y, -x, z), (-x, -y, z+1/2), (y, -x+y, z+1/2), …… (x, x-y, z), (y, x, z+1/2), (x-y, -y, z+1/2), (-x, -x+y, z+1/2) \\

        \cmidrule[0.6pt]{1-5}
        166 & Trigonal & 1036 & 36 & (x, y, z), (-y, x-y, z), (-x+y, -x, z), (y, x, -z), (x-y, -y, -z), …… (y+1/3, -x+y+2/3, -z+2/3), (x-y+1/3, x+2/3, -z+2/3), (-y+1/3, -x+2/3, z+2/3), (-x+y+1/3, y+2/3, z+2/3), (x+1/3, x-y+2/3, z+2/3) \\

        \cmidrule[0.6pt]{1-5}
        139 & Tetragonal & 1599 & 32 & (x, y, z), (-x, -y, z), (-y, x, z), (y, -x, z), (-x, y, -z), (x, -y, -z), …… (x+1/2, -y+1/2, z+1/2), (-x+1/2, y+1/2, z+1/2), (-y+1/2, -x+1/2, z+1/2), (y+1/2, x+1/2, z+1/2) \\

        \cmidrule[0.6pt]{1-5}
        62 & Orthorhombic & 1429 & 8 & (x, y, z), (-x+1/2, -y, z+1/2), (-x, y+1/2, -z), (x+1/2, -y+1/2, -z+1/2), (-x, -y, -z), (x+1/2, y, -z+1/2), (x, -y+1/2, z), (-x+1/2, y+1/2, z+1/2) \\

        \cmidrule[0.6pt]{1-5}
        12 & Monoclinic & 1429 & 8 & (x, y, z), (-x, y, -z), (-x, -y, -z), (x, -y, z), (x+1/2, y+1/2, z), (-x+1/2, y+1/2, -z), (-x+1/2, -y+1/2, -z), (x+1/2, -y+1/2, z) \\

        \cmidrule[0.6pt]{1-5}
        2 & Triclinic & 1230 & 2 & (x, y, z), (-x, -y, -z) \\
        \bottomrule[0.8pt]
        \bottomrule[0.8pt]     
    \end{tabular}}
    \vspace{-2mm}
\end{table*}

\newpage
\begin{figure}[H]
    \centering
    \includegraphics[width=16cm]{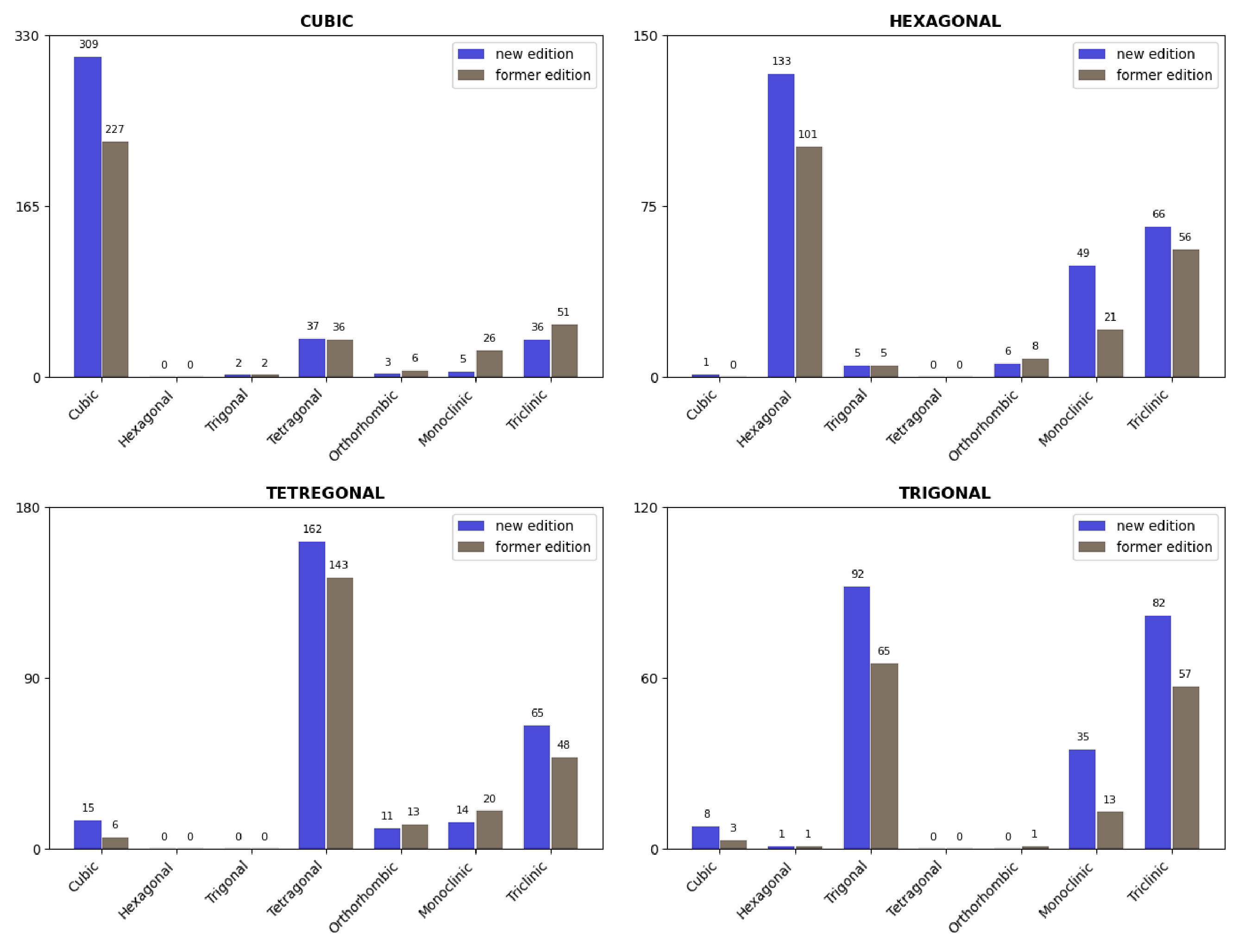}
    \caption*{\textbf{Figure S1:} The crystal system distribution of the crystals sampled in symmetry-based generation task(Figure3), with random space group chosen in dataset for each type of system. This experiment is aimed to showcase the robustness of SLICES-PLUS. Blue bars refer to the valid samples represented in SLICES-PLUS, while brown bars refer to those represented in SLICES.}
    \label{fig:fig7}
\end{figure}
\newpage
\begin{figure}[H]
    \centering
    \includegraphics[width=16cm]{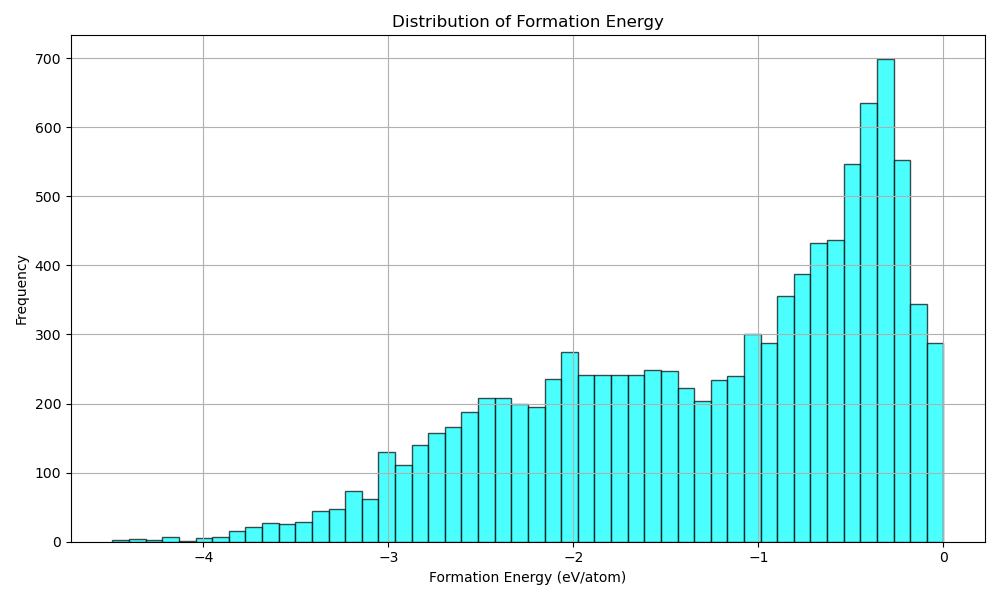}
    \caption*{\textbf{Figure S2:} Distribution of formation in the dataset(Figure4).}
    \label{fig:fig8}
\end{figure}
\newpage
\begin{figure}[H]
    \centering
    \includegraphics[width=16cm]{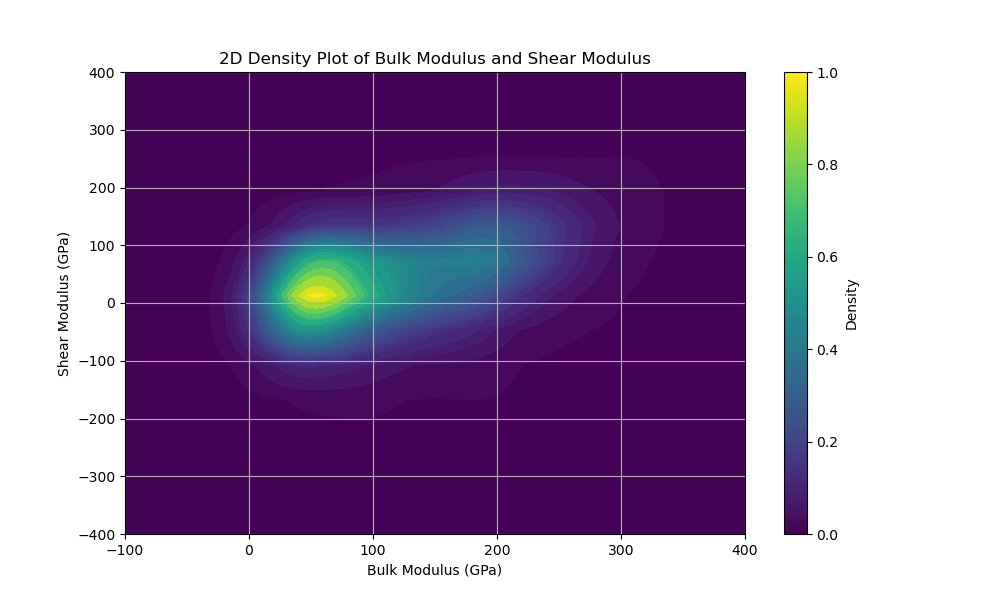}
    \caption*{\textbf{Figure S3:} Distribution of bulk-shear modulus in the dataset(Figure5).}
    \label{fig:fig9}
\end{figure}

\end{document}